# Nonlinear FM Waveform Design to Reduction of sidelobe level in Autocorrelation Function

Roohollah Ghavamirad
Department of Electrical Engineering
K. N. Toosi University of Technology
Email: r_ghavamirad@ee.kntu.ac.ir

Hossein Babashah
Department of Electrical Engineering
Sharif University of Technology
(Email: babashah_hossein@ee.sharif.edu)

Mohammad Ali Sebt
Department of Electrical Engineering
K. N. Toosi University of Technology
Email: sebt@kntu.ac.ir

*Abstract*— **This paper will design non-linear frequency modulation (NLFM) signal for Chebyshev, Kaiser, Taylor, and raised-cosine power spectral densities (PSDs). Then, the variation of peak sidelobe level with regard to mainlobe width for these four different window functions are analyzed. It has been demonstrated that reduction of sidelobe level in NLFM signal can lead to increase in mainlobe width of autocorrelation function. Furthermore, the results of power spectral density obtained from the simulation and the desired PSD are compared. Finally, error percentage between simulated PSD and desired PSD for different peak sidelobe level are illustrated. The stationary phase concept is the possible source for this error.**

## I. Introduction

The principal advantages of pulse compression with NLFM waveform are high Signal-to-Noise Ratio, easily implementable transmitter due to constant envelope of the transmitted signal, and more reduction of sidelobe level in autocorrelation function (ACF) with regard to linear frequency modulation (LFM) method [1-7].

In this paper, we analyze variation of peak sidelobe level with regard to mainlobe width for four different window functions. In order to reduce sidelobe level of autocorrelation function, shaping the power spectral density can be used, since autocorrelation function is the inverse Fourier transform of the PSD [8], [9]. To achieve this goal, Chebyshev, Kaiser, Taylor, and raised-cosine windows are considered as four desired PSDs. Afterwards, group time delay function in frequency domain is carried out using the stationary phase concept (SPC). This was done by integration of PSD function and applying boundary conditions [10]. Then group time delay inverse function is computed yielding frequency function in time domain. The results of autocorrelation function obtained from the simulation are presented for all the aforementioned window functions and comparison between sidelobe level and mainlobe width are made.

As the sidelobe level reduces, the mainlobe width increases [11], [12]. This can diminish target detection accuracy. In order to preserve the accuracy, an optimized condition should be considered which is discussed in the paper. Furthermore, the resulting PSD plot and desired PSD plot were compared in order to calculate error percentage. It should be mentioned that the SPC is the possible source for this error due in large to the fact that SPC is based on approximated relations.

The remainder of the paper is organized as follows. NLFM signal is designed for four different windows in section II. Section III discusses the simulation results; Section IV concludes the paper.

## II. NLFM Waveform Design

### A. Method based on Stationary Phase Concept

The first study on using SPC for NLFM was carried out in 1964 by Fowle [8]. To design NLFM signal, a low-pass signal denoted as $x(t)$ is considered.

$$x(t) = a(t)\exp(j\varphi(t)) \quad (1)$$

Where $a(t)$ and $\varphi(t)$ are amplitude and phase of the signal respectively. The instantaneous frequency $f_m$ at time $t_m$ is defined as time derivative of the phase:

$$f_m = \frac{1}{2\pi}\frac{d\varphi(t_m)}{dt} \quad (2)$$

We shall write the Fourier transform of $x(t)$ as $X(f)$. Based on SPC, PSD at a frequency is large if the rate of frequency change at that time is relatively small. Thus, the relation between PSD and frequency variation can be expressed as:

$$|X(f_m)|^2 \approx 2\pi\frac{a^2(t_m)}{|\varphi''(t_m)|} \quad (3)$$

According to (3) it is clear that PSD is inversely proportional to second time derivative of phase and it is directly proportional with signal amplitude. Since frequency changes with regard to time is linear for LFM, the second time derivative of phase is constant and the PSD depends on signal amplitude. In NLFM approach, signal amplitude is assumed to be constant ($a(t) = A$, A is constant) and PSD is depended on the value of second order phase derivative. Let us consider $X^2(f)$ with an desired PSD such as $Z^2(f)$, where $Z^2(f)$ only depends on the second derivative of phase of $x(t)$ signal. Equation (3) is reproduced in frequency domain as:

$$\Phi''(f) \approx \frac{Z^2(f)}{\frac{A}{2\pi}} = kZ^2(f) \ , \ k = constant \quad (4)$$

If $Z(f)$ is defined in the frequency range of $-B/2 \leq f \leq B/2$ ($B$ is the bandwidth), then the first derivative of phase $\Phi'(f)$ is resulted by integral of the second derivative $\Phi''(f)$ which can be expressed as:

$$\Phi'(f) = \int_{-\frac{B}{2}}^{f} \Phi''(\theta)d\theta = k\int_{-\frac{B}{2}}^{f} Z^2(\theta)d\theta \quad (5)$$

The following relation defines group time delay function $T_g(f)$ as:

$$T_g(f) = -\frac{1}{2\pi}\Phi'(f) \quad (6)$$

Substituting of (6) in (5) yields:

$$T_g(f) = -\frac{k}{2\pi}\int_{-\frac{B}{2}}^{f} Z^2(\theta)d\theta = k_1\int_{-\frac{B}{2}}^{f} Z^2(\theta)d\theta + k_2 \quad (7)$$

Where $k_1$ and $k_2$ are the integration constant which are determined using the boundary conditions of group time delay function. If the signal pulse width is equal to T, then the boundary conditions of group time delay function written in the form $T_g(B/2) = T/2$ , $T_g(-B/2) = -T/2$ [10]. Moreover, instantaneous frequency as a function of time is given by

$$f(t) = T_g^{-1}(f) \quad (8)$$

Finally, the phase of the designed signal can be obtained from frequency function using the following equation.

$$\varphi(t) = 2\pi \int_{-\frac{T}{2}}^{t} f(\theta)d\theta \quad (9)$$

### B. Different Desired PSDs Analysis

#### 1) Raised Cosine Window

NLFM waveforms can be designed using the raised cosine window. For reason of better results, second order raised cosine window is considered in this paper. Thus, PSD relation can be written in the following form

$$w(n) = k + (1-k)cos^2\left(\frac{\pi n}{M-1}\right), \quad |n| \leq \frac{M-1}{2} \quad (10)$$

Where $k$ is a constant value which can control sidelobe level and M is window length. Assume that $f = nB/(M-1)$ then, by integration and applying the boundary condition, group time delay function can be derived as:

$$T_g(f) = T\left(\frac{f}{B} + \left(\frac{1-k}{1+k}\right)\frac{\sin\left(\frac{2\pi f}{B}\right)}{2\pi}\right) \quad (11)$$

The inverse function of $T_g(f)$ complicated and it does not have closed-form expression. To illustrate the result, a simulation of inverse function for $T_g(f)$ was performed.

#### 2) Chebyshev Window

The optimal Dolph-Chebyshev window transform in closed-form is given by:

$$W(m) = \frac{\cos\{M\cos^{-1}[\beta\cos(\frac{\pi m}{M})]\}}{\cosh[M\cosh^{-1}(\beta)]}, \quad (12)$$

$$m = 0,1,2,\ldots,M-1$$

$$\beta = \cosh\left[\frac{1}{M}\cosh^{-1}(10^\alpha)\right] \ , \ \alpha \approx 2,3,4 \quad (13)$$

Where M and $\alpha$ are window length and parameter for determining the Chebyshev norm of the sidelobes to be $-20\alpha$ dB, respectively. Afterwards, zero-phase Dolph-Chebyshev window was calculated using the inverse DFT of $W(m)$ [13].

$$w(n) = \frac{1}{N}\sum_{m=0}^{M-1} W(m).\exp\left(\frac{j2\pi mn}{M}\right), \quad |n| \leq \frac{M-1}{2} \quad (14)$$

#### 3) Kaiser Window

The Kaiser window can be computed through approximation of the discrete prolate spheroidal sequence (DPSS) window using Bessel functions which is written in the following form:

$$w(n) \triangleq \begin{cases} \frac{I_0\left(\pi\alpha\sqrt{1-\left(\frac{n}{M/2}\right)^2}\right)}{I_0(\pi\alpha)} & |n| \leq \frac{M-1}{2} \\ 0 & elsewhere \end{cases} \quad (15)$$

In the above equation, $I_0$ and $\alpha$ represent the zero-order modified Bessel function of the first kind and an arbitrary real number, respectively. In fact, $\alpha$ shows the shape of the window in the frequency domain and parameter $\beta \triangleq \pi\alpha$. M is the length of window [13]. The group time delay function in the Chebyshev and Kaiser windows should be determined numerically.

#### 4) Taylor Window

The equation that express Taylor weighting function is as follows:

$$w(n) = 1 + \sum_{m=1}^{\bar{n}-1} F_m \cos\left(\frac{2\pi mn}{M-1}\right), \quad |n| \leq \frac{M-1}{2} \quad (16)$$

In (16), $F_m = F(m,\bar{n},\eta)$ is Taylor coefficients of the mth order. $\eta$ and $\bar{n}$ show ratio of mainlobe over sidelobe level and number of sidelobes at equal level, respectively. M also identifies window length [4]. Assume that $f = nB/(M-1)$, by integration and applying the boundary condition, group time

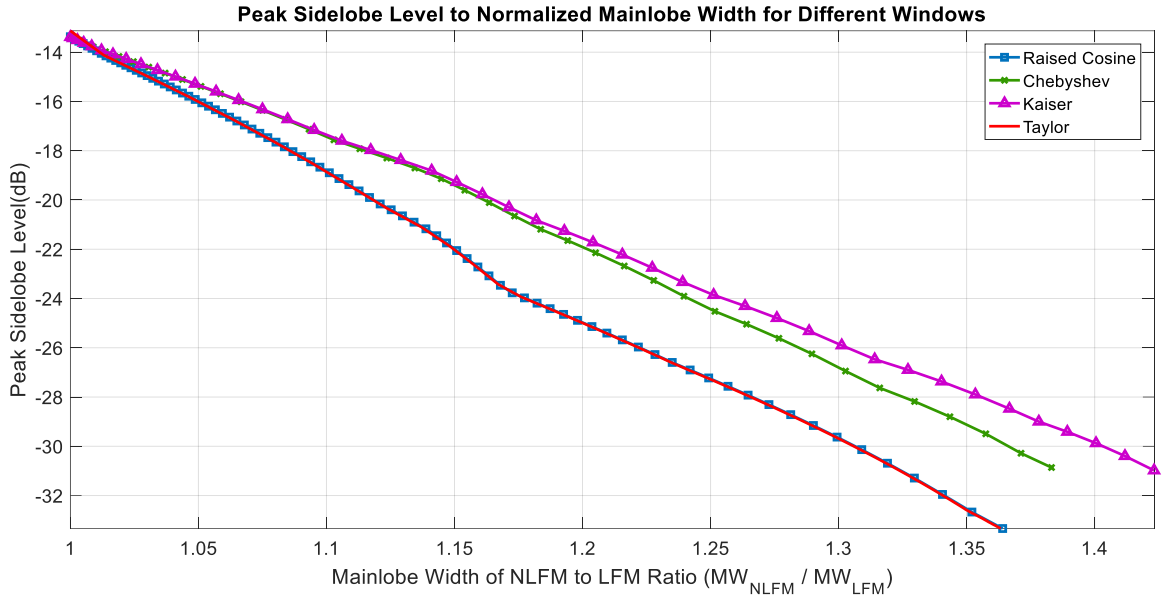

Figure 1. Changes in peak sidelobe level with regard to normalized mainlobe width in autocorrelation function for four different windows.

delay function in Taylor window can be derived as:

$$T_g(f) = T\left(\frac{f}{B} + \frac{1}{2\pi}\sum_{m=1}^{\bar{n}-1}\frac{F_m}{m}\sin\left(\frac{2\pi m f}{B}\right)\right) \quad (17)$$

In this paper $\bar{n}$ is assumed to be 2, so group time delay function is then given by

$$T_g(f) = T\left(\frac{f}{B} + \frac{F_1}{2\pi}\sin\left(\frac{2\pi f}{B}\right)\right) \quad (18)$$

The inverse function of $T_g(f)$ in the Taylor window should be carried out numerically.

## III. SIMULATION AND RESULTS

Table 1 lists the data that was used in simulation. Comparison between changes in mainlobe width and peak sidelobe level (PSL) are made. Mainlobe width is considered in the order of -3 dB then, PSL can be computed by the following equation [14]

$$PSL(dB) = 20\log_{10}\left(\frac{\max(|R(\tau)|)}{|R(0)|}\right), \quad z_1 \leq |\tau| \leq \frac{T}{2} \quad (19)$$

Where $|R(\tau)|$ is the amplitude of autocorrelation function of NLFM signal in the time delay of $\tau$ second, and $z_1$ is the first-null of $|R(\tau)|$. Figure 1 illustrates changes in PSL versus normalized mainlobe of autocorrelation function of NLFM signal for four different windows. Normalized mainlobe width is calculated using mainlobe width of autocorrelation NLFM signal divide by LFM signal.

TABLE I. SIMULATION PARAMETERS

| Parameter | Value | Unit |
|---|---|---|
| Pulse width | 2.5 | μs |
| Bandwidth | 100 | MHz |
| Sampling rate | 1 | GHz |

The results obtained for raised cosine and first-order Taylor windows bear a close resemblance due to their similar group time delay function. At the point where the normalized mainlobe width is equal to 1.364, peak sidelobe level is -33.34 dB for raised cosine and first-order Taylor indicating better condition with regard to Chebyshev and Kaiser cases. Figure 2 illustrates normalized NLFM and LFM autocorrelation function amplitude signal for four aforementioned windows. From this figure it can be seen that every window experience maximum sidelobe level reduction. A comparison of the two results reveals that decrease in NLFM sidelobe level is favorably more significant than LFM. As mentioned earlier, raised cosine and Taylor autocorrelation function are similar to one another.

The results of PSD obtained from the simulation and the desired PSD for different windows can be compared in Figure 3. The observed difference between simulated PSD and desired PSD can be explained by the fact that SPC is based on approximated relations. To reduce error percentage, error optimization methods can be used. Figure 4 depicts error percentage of the simulated PSD and desired PSD against peak sidelobe level of autocorrelation function for NLFM signal. It is apparent from this figure that the reduction of sidelobe level results in lower calculated error percentage.

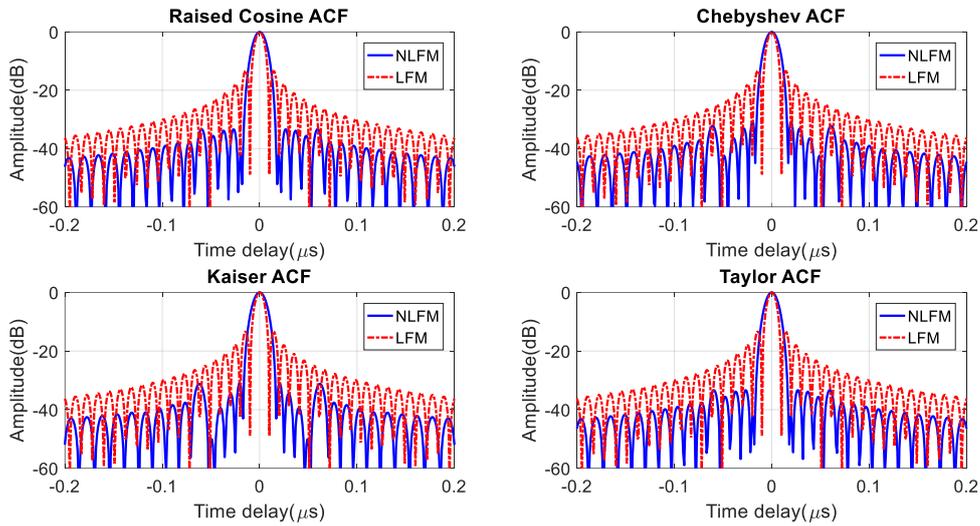

Figure 2. NLFM and LFM signal autocorrelation function amplitude for four different window functions.

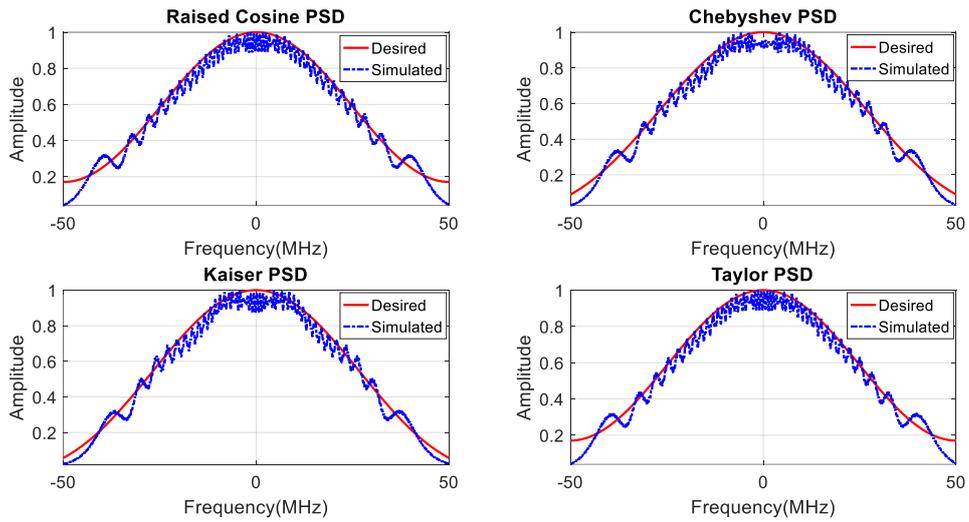

Figure 3. Ampitude of desired PSD and simulated PSD for four different window functions.

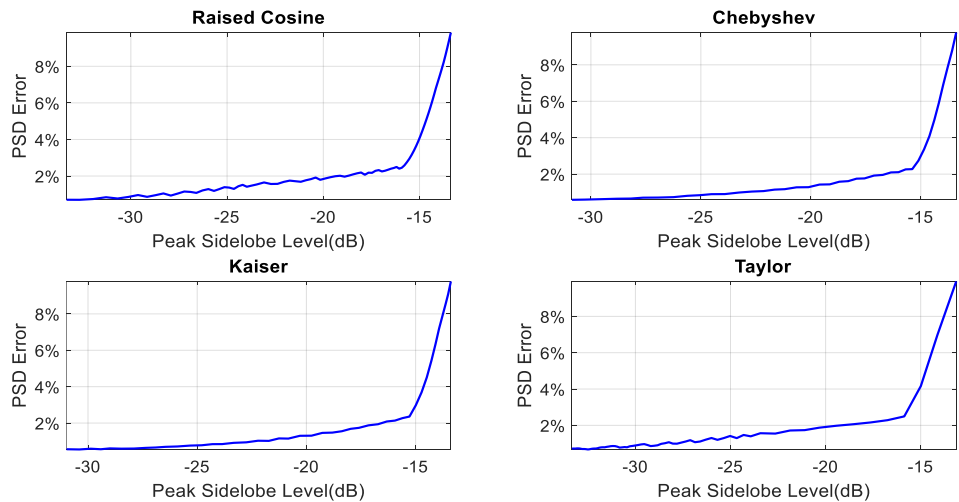

Figure 4. Error percentage of the simulated PSD and desired PSD against peak sidelobe level of autocorrelation function for designed NLFM signal.

IV. CONCLUSION

Based on the results, it can be concluded that reduction of sidelobe level in NLFM signal can lead to increase in mainlobe width of autocorrelation function, but this increment is not significant with regard to sidelobe level. The results obtained also indicate that raised cosine and first-order Taylor windows perform better than Chebyshev and Kaiser windows on reduction of sidelobe level. Approximated relations of stationary phase concept result in low error percentage.